\def\grs{GRS~$1915$+$105$}
\def\X1859{XTE~J$1859$+$226$}

\def\rxte{{\it{RXTE}}}
\def\chisq{$\chi^2_\nu$}

\documentclass[]{aastex}
\usepackage{emulateapj5}

\usepackage{amssymb}
\usepackage{natbib}
\usepackage{epsfig}
\usepackage{epsf}
\usepackage{color}
\definecolor{red}{rgb}{0.7,0,0}
\definecolor{blue}{rgb}{0,0,0.7}
\def\correc#1{{#1}}

\shorttitle{The cathedral QPO in XTE J1859+226}
\shortauthors{Rodriguez \& Varni\`ere}
\begin{document}
\title{The puzzling harmonic behavior of the Cathedral QPO in XTE J1859+226}

\author{J. Rodriguez\altaffilmark{1} \& P. Varni\`ere\altaffilmark{2}}

\altaffiltext{1}{Laboratoire AIM, UMR 7158, CEA/DSM - CNRS - Universit\'e Paris Diderot, IRFU/SAp, F-91191 Gif-sur-Yvette, France}
\altaffiltext{2}{Laboratoire APC, UMR 7164, CNRS-Universit\'e Paris Diderot-CEA/DSM, 10 rue Alice Domon et Leonie Duquet,
75205 Paris Cedex 13, France. }

\begin{abstract}We present a spectral and temporal analysis of the Cathedral QPO detected  
in the power density spectra of the black hole binary and microquasar \X1859 obtained with \rxte. This peculiar type of 
QPO has been seen on two occasions (on MJDs 51574.43 and 51575.43)  during the 1998 outburst of this source. 
It manifests as two peaks with similar amplitudes ($\sim3\%$ and $\sim5\%$ RMS) and harmonically related centroid frequencies ($\sim 3$ and 
$\sim 6$ Hz). The temporal properties of these two peaks are different: the amplitude of the $\sim3$~Hz feature varies, in anticorrelation 
with the count rate, by about  $\sim50\%$. The $\sim6$~Hz feature, on the other hand, shows a  slight increase ($\sim 7\%$) of its amplitude with 
count rate. The RMS-spectra of the two peaks are also quite different. The $\sim3$~Hz feature is softer than the other one,
and although its RMS amplitude increases with energy it shows a cut-off at an energy of $\sim 6$ keV. The RMS  of the 6~Hz increases up 
to at least 20~keV. We also study the  bicoherence, $b^2(\mu,\nu)$ of both observations. At the diagonal position 
of the peaks the values $b^2(\sim3,\sim3)$ and  $b^2(\sim6,\sim6)$ are rather high and similar to what has been reported in the case 
of the type C QPOs of GRS 1915+105. By comparison with the latter source the fact that the bicoherence of the $\sim 3$~Hz feature 
is higher than than of the other peak, would tend to indicate that the $\sim 3$~Hz is the fundamental QPO, and the other its first harmonic.
The value of $b^2(\sim3,\sim6)$  is, however, low and therefore indicates a behavior that is different than that seen in GRS 1915+105.
We discuss the implications of these differences in the context of an harmonic relationship of the peaks, and 
suggest that, rather than pure harmonics, we may see different modes of the same underlying phenomenon competing to produce
QPOs at different frequencies.
\end{abstract}

\section{Introduction}
Black Hole Binaries \citep[BHB, e.g.][for a review]{remillard06} transit through different  'spectral states' during their
outbursts.  These are defined both by the spectral and temporal parameters respectively obtained through analyses of 
 their energy spectra and power density spectra (PDS). In the soft state, where the  emission is dominated by a bright 
and warm ($\sim 1$ keV) accretion disk, the level of variability is weak, and the PDS is power law like. 
On the other hand, in the hard state, when the disk is much colder ($\leq 0.5$ keV) and thought to be truncated 
at a large distance from the accretor, the level of variability is much higher and shows a band limited noise component. 
 Other states exist and can be seen as intermediate between the Hard and Soft ones. 
 See e.g. \citet{remillard06,homan05b} for recent reviews and precise classifications of the BHB's states. 
 Furthermore, 
quasi-periodic oscillations in the 'low frequency' range (0.1--20 Hz, hereafter referred to as LFQPO) are seen  in the 
hard and during the intermediate states. These LFQPOs have been further classified into  types A, B, or C 
based on their typical frequencies,  total RMS amplitude, and time lags \citep[e.g.][]{remillard02, casella05}.  
We have recently proposed  a tentative classification of states based on the presence of the 
different types of QPOs \citep{varniere11}. \\
\indent The exact origin of QPOs is still a matter of debate: they could be due to global
oscillations of the disk \citep[e.g.][and ref. therein]{wagoner01}, Lense-Thirring precession \citep[e.g.][and ref. therein]{ingram09}, oscillating shocks 
 \citep[e.g.][and ref. therein]{molteni96}, or MHD instability \citep{tagger99}. From the observational side it seems clear that the inner disk somehow 
 sets the frequency of LFQPOs  \citep[e.g.][]{muno99,rodrigue02_qpo,rodrigue02_aei, rodrigue04_1550, Mikles09}.  LFQPOs are, however, strong when 
 a strong hard component is seen in the energy spectra and their frequency is  correlated with the power law photon 
 index \citep[e.g.][]{vignarca03, shaposhnikov07}. This last point would tend to indicate an origin in relation to the corona. 
 The amplitude of QPOs increases with the energy up to a cut-off whose energy is  variable  \citep{rodrigue04_1915,rodrigue08_1915b}. \\
\indent These LFQPOs usually manifest in the PDSs as  powerful peak referred to as the fundamental and a series of (sub) harmonics
usually of much fainter amplitudes.  Very little focus has been put so far on these harmonics, even if it is clear that they do not completely share 
the same properties as the fundamental. They can show different signs for their time lags, as in the case of 
type B QPOs \citep{casella04}, and show different shapes of the  RMS spectra \citep{cui99,homan01,rao10}. The true identification of the 
fundamental has recently been questioned \citep[e.g.][]{rao10}, which raises questions related to the genuineness of the harmonic
relationship of the peaks.  \\

\indent \X1859 was discovered on 1999 October 9 with the \rxte\ All Sky Monitor \citep{wood99} as it was entering into outburst. 
It is a microquasar given the observations of relativistic ejections in radio \citep{brocksopp02}, and the observations of  low and high 
frequency QPOs led \citet{cui00} to classify it as a candidate BHB. An extensive timing analysis of this source is presented by 
\citet{casella04}. Like XTE J1550$-$564, \X1859\ displays all three types of  LFQPOs.
In two particular observations, 
  during the 1999 outburst, \citet{casella04} observe the presence of two peaks with harmonically related frequencies,  but unlike 
any other cases, similar RMS-amplitudes. These are referred to as the 'Cathedral' QPO \citep{casella04}: 
the strongest peak (and highest in frequency)  has hard lags (the hard X-ray lag
behind the soft X-rays) and is interpreted as the fundamental peak, while  the lowest frequency peak has soft lags and is  the 
sub-harmonic. Interestingly these two observations, although separated by about a day, are also separated by an observation 
showing another type of QPO indicative of a spectral transition \citep{casella04}. \\
\indent  The existence of this Cathedral QPOs showing two harmonically related peaks of similar amplitude  raises challenging 
questions for all theoretical models. To obtain a clearer view of the properties of these peaks we performed a complete study of the 
QPO structure from these two observations, and compare the temporal and spectral behaviors of the two peaks. The organization 
of the letter is as follows: we start presenting the observations ID and data reduction methods. We then describe our results in Section 3, starting 
with the temporal evolution of the source, the fit to the broad band PDS,  the energy dependences of the QPOs, and we end this part 
by presenting the bicoherence of the observations. We  discuss our findings in the last section of this letter.

\section{Observations and data reduction}
We focus on \rxte\ observations  40124-01-24-00 (Obs.~1) and 40124-01-27-00 (Obs.~2) respectively 
made on 1999 october, 23 (MJD 51474.43) and 24 (MJD 51475.43), near the peak of the 1999 outburst. 
The full process of \rxte/PCA data reduction was made with the {\tt{HEASOFT v6.9}} software package.
We reduced the Binned and Event mode data following standard procedures \citep[see e.g.][]{rodrigue08_1915b}
to obtain light curves filtered from low elevation above the Earth, large offset from the source, and PCU breakdown.
We extracted $7.8125\times 10^{-3}$~s binned light curves in several energy ranges used for the fine timing analysis
, and a  $\sim2$--$15$~keV light curve from the Binned data with a time bin of 8 s (Fig. ~\ref{fig:dynpo}) to characterize the overall 
behavior of the source over the observations. Note that this range contains most of the counts emitted by the source.
Power density spectra (PDS) were then produced with {\tt{Powspec v1.0}} on intervals of $16$~s in the range
$0.0625$--$64$~Hz, all intervals being further averaged together before the fitting process. A dynamical PDS (DPDS)
was also computed between 0.25 and 64 Hz, to study the variations of the QPOs with time (Fig. \ref{fig:dynpo}).
The PDSs were fitted between  $0.0625$ and $40$~Hz in {\tt{XSPEC v12.6.0}}. The background rate was taken 
into account when estimating the RMS amplitudes of the different features following $A_{net}=A_{raw}\times\frac{S+B}{S}$, 
with $A$ the amplitude, $S$ is the source net rate, and $B$ the background rate \citep{berger94,rodrigue04_1915,rodrigue08_1915b}.\\

\section{Results}
\subsection{Time evolution of the source: light curve and dynamical PDS}
\label{sec:dynpo}
\begin{figure}
\epsscale{0.5}
\caption{Upper panel: 2--40 keV dynamical power spectrum of \X1859. Lower panel: 2--40 keV light curve.}
\plotone{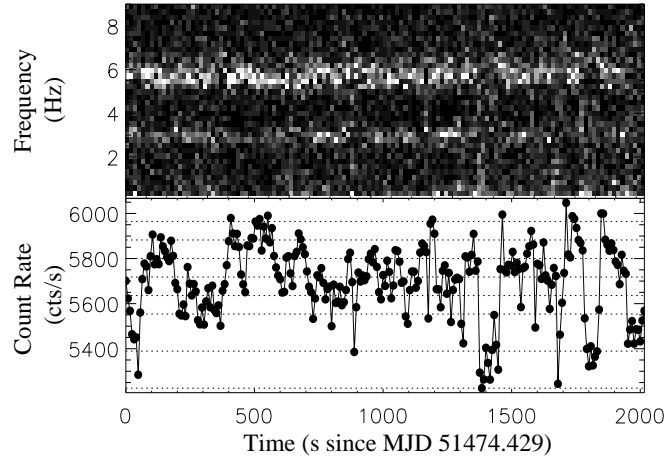}
\label{fig:dynpo}
\end{figure}

The 2--15 keV \X1859\ PCA light curve binned at 8~s and DPDS of Obs. 1 are reported in Fig.~\ref{fig:dynpo}. Large variations  around a mean 
raw (net) count rate of 5700 cts/s (5676 cts/s) are clearly visible. We especially note the presence of two rather broad dips lasting respectively  
$\sim100$~s and $\sim50$~s  near relative times 1350~s and 1775~s, and a third occurring near the end of the observation (at t$\sim2050$~s).
The DPDS shows the presence of two strong features (relatively to the overall noise) around 3 and 6~Hz. These two features are quite thin
and indicate the presence of QPOs at these frequencies. Interestingly the two QPOs seem to have different behavior with time  (Fig.~\ref{fig:dynpo}):  
the QPO with the smallest frequency is, on average, much weaker 
than the other one, and is strong only when the count rate is around its mean value. It is, in particular quite weak during the small
flares, and is not visible during the three dips. The highest frequency QPO, on the other hand, seems, in term of 
power, more stable and seems to vary significantly only during the dips (Fig.~\ref{fig:dynpo}). The feature seems rather broad around $\sim 6$~Hz, which 
may indicate some rapid variations of the frequency, or simply an intrinsically low coherence  QPO. A very similar behavior (not shown) is also seen in both 
the light curve and DPDS of Obs. 2.\\
\subsection{Broad band PDS}
\label{sec:bbPDS}
We started with fitting the large band PDSs. We did not subtract the white noise and preferred adding a constant
to our fit model to account for this component. Following \citet{casella04} we fitted the continuum with the 
sum of three broad and three narrow  Lorentzians (Fig. \ref{fig:PDS}) on top of the white noise component. 
This model yields a good fit with \chisq=1.13 (resp. 1.09) for 118 degrees of freedom (DOF) for Obs.~1 (resp. Obs.~2). 
The parameters of the three thin peaks are reported in Table~\ref{tab:qpofit} for both observations. The 
parameters of all three peaks are compatible between the two observations, which lends credence to their complete similarity.
In the remainder of the paper only the results of Obs.~1 are precisely described and presented in the figures. Note that in all 
cases the same analysis was performed on Obs.~2 and the results and trends observed are consistent with those of Obs.~1.
The third peak is   compatible with $4\times \nu_1$, and $2\times \nu_2$ at the $\sim 2\sigma$ level, and  is likely to be an harmonic 
of one of the two main peaks. In the remainder of this paper, we will refer to either peak 1, 2, or 3, or QPO 1, 2, or 3 
for the peaks at $\sim2.9$, $\sim5.8$, and $\sim 11.2$ Hz respectively.\\
\indent Although the fit statistic  is quite good, we remark some  residuals on the lower shoulder of QPO$_2$.  This effect is also
mentioned by \citet{casella04} in Obs. 2, and these authors added a Gaussian to better represent the peak. 
In our case, the addition of another thin Lorentzian at $\sim5.5$~Hz improves the fit to \chisq=0.90 for 137 DOF,
and corrects the defect previously present in the residuals. The feature is, however, poorly defined, and its parameters 
are badly constrained. We verify, by re-doing the whole analysis  that it had no significant impact on the 
other peaks, and since no influence was found it was omitted from our study, and is not further discussed here. \\
\indent As mentioned in the previous section, QPO$_1$ seems more intermittent that the second peak (Fig.~\ref{fig:dynpo}). 
In order to quantify and study any possible dependence of the QPOs amplitudes with the count rate (Sec. \ref{sec:dynpo}),  
we followed a procedure similar  to that presented in \citet{heil11}. Each observation was separated into ten count rate intervals of 
equal width. Each interval therefore has a width equal to $\frac{Max(CR)-Min(CR)}{10}$~cts/s. Due to short accumulation times, and thus 
poor statistical quality of the resulting PDSs, some of pairs of intervals were combined together. In the case of Obs.~1 this rebin concerned 
intervals 1\&2, and intervals 3\&4. The final division of Obs.~1 is represented in Fig.~\ref{fig:dynpo}. 
We then extracted a $7.8125\times 10^{-3}$~s binned light curve per time interval of 8~s, and produced a PDS from each of these light curves.  
Individual PDSs belonging to the same count rate interval were averaged together to produce the final count rate dependent PDSs.
The latter were then fitted to estimate the parameters of the peaks, with a special focus on their rms amplitudes. No specific trend is seen between 
the frequency or the coherence of any of the two peaks with the count rate.  \correc{In order to ease the comparison with the QPOs of 
XTE~J1550$-$564 \citep{heil11}, we represented the evolution of the absolute RMSs (in terms of cts/s)} of the two peaks with the count rate 
 in Fig.~\ref{fig:QPOvsCount} for the particular case of Obs.~1. 
The amplitude of QPO$_1$ decreases with increasing count rate, while that of QPO$_2$  increases slightly  (Fig.~\ref{fig:QPOvsCount}) 
\correc{and show a possible linear trend.} 
Between the first real detections of the two peaks  (in the second count rate bin) and the last bin, 
QPO$_1$ varies from an RMS amplitude of $2.7\pm0.3\%$ to $1.35\pm0.34$ (a variation of $50\%$ in amplitude) while QPO$_2$ varies  from 
$3.6\pm0.4\%$ to $3.9\pm0.2\%$ (a non-significant variation of $\sim 7.7\%$ in amplitude). 


\subsection{Energy dependence of the QPOs}
 To study the energetic dependence of the QPOs and produce the RMS-Spectra (Fig.~\ref{fig:QPOspec}), we fitted each 
 energy dependent PDS with the statistically required number of broad features to account for the continuum. We remark here 
 that these features also have complex energetic dependences, and thus the different fits either require 1,2, or 3 broad Lorentzians. 
 The study of these is, however, beyond the scope of this letter and we will not discuss them further. 
 We then added the thin Lorentzians to account for the QPO. The third peak is most of time undetectable and no spectrum can be acquired
 for it. The resultant RMS-Spectra   for QPO$_1$ and QPO$_2$ from Obs. 1 are reported in  Fig.~\ref{fig:QPOspec}. An almost identical behavior 
 is seen in the QPO spectra obtained from Obs. 2 (not shown). \\
\indent The two QPOs  share a common trend:  their amplitude first increases with energy
before reaching a plateau. Such QPO-spectra are common for all types (A, B, C) of LFQPOs in this source \citep{casella04}, and have, for example, also
been seen in \grs\ \citep{rodrigue04_1915,rodrigue08_1915b}, and XTE J1550$-$564 \citep{homan01,rodrigue04_1550}. 
The fact that the normalization of the spectra is different is not unexpected since 
the two peaks have different total amplitudes. The precise shape and typical parameters (energy of 
the break, slope, ...) are, however,  clearly different (Fig.~\ref{fig:QPOspec}). This is illustrated in the right panel of Fig.~\ref{fig:QPOspec}, where we normalized 
the RMS spectrum of QPO$_2$ by that of QPO$_1$. This representation also clearly shows that QPO$_2$ has a steeper (harder) spectrum than QPO$_1$. 
The latter first increases up to $\sim5.7$ keV and is then flat
until $\sim 20$~keV. QPO$_2$ is undetectable in the first energy bin (and is thus fainter than QPO$_1$). It 
increases up to $\sim20$~keV where its plateau is reached. 

\subsubsection{Bicoherence of the Cathedral QPO}
\label{sec:bicoh}
The bispectrum and the related bicoherence permit any possible coupling between the different components of the PDS to be 
studied, and thus allow one to go beyond the diagnosis brought by the PDSs 
\citep[see e.g.][and references therein for a deeper discussion on these aspects]{maccarone11}.
Dividing the light curve in K segments of equal length, the bicoherence is expressed as:  
$$b^2(\nu,\mu)=\frac{|\sum_{i=0}^{K-1} X_i(\nu)X_i(\mu)X_i(\nu+\mu)|^2}{\sum_{i=0}^{K-1} |X_i(\nu)X_i(\mu)|^2 \times \sum_{i=0}^{K-1}|X_i(\mu+\nu)|^2}$$
where $X_i(\nu)$ is the $\nu$ component of the discrete Fourier transform of the $i^{th}$ interval 
\citep[see e.g.][for the description, restriction, interpretation and 
applications to various astrophysical sources, including BHBs]{maccarone02,uttley05,maccarone11}. 
We calculated the 2--15 keV bicoherences of \X1859\ between $6.25\times10^{-2}$~Hz and $16$~Hz, with a frequency 
resolution of $0.0625$~Hz.   Fig.~\ref{fig:bicoh} shows a zoom on the  bicoherence plot we obtained in the case of Obs. 1 
(the plots and results obtained in the case of Obs. 2 are almost identical). 
Note that by definition the bicoherence is symmetric with respect to the first diagonal ($b^2(\nu,\mu)=b^2 (\mu,\nu)$). 
The highest values of the bicoherence are reached in a region around ($\nu_1,\nu_1$) (the latter being indicated by the white box close 
to the bottom left corner in Fig.~\ref{fig:bicoh}), between $\sim2.8$ and $\sim3.2$~Hz 
along the first diagonal (Fig.~\ref{fig:bicoh}), with a slight broadening at low frequencies. 
A local maximum is also reached around the frequency of QPO$_2$ (illustrated by the second white box in Fig.~\ref{fig:bicoh}). Note that 
the mean value of $b^2$ calculated in the white regions represented in Fig.~1 is higher around QPO$_1$ than around QPO$_2$.   
 The mean value of $b^2$ over the $0.0625$--$16$ Hz range is $0.008$ with a an RMS of about $0.008$. The value of $b^2$ around 
 $(\nu_2,\nu_1)$ (black box in Fig.~\ref{fig:bicoh})  is quite low. At the position of the two peaks we obtain $b^2(\nu_2,\nu_1)\sim0.01$, 
 a value within the statistical fluctuations around the mean of  $b^2$, and clearly compatible with "noise". Note 
 that the same results are obtained when looking at the mean value of $b^2$ in broad (e.g. $5\times5$ frequency-pixels) regions centered
on  $(\nu_2,\nu_1)$. Overall, this tends to show that either the two peaks are not coupled
 or that their coupling is very weak. 

\section{Discussion and conclusions}
\begin{figure}
\epsscale{0.5}
\caption{ 2--40 keV power spectra showing the presence of the 'cathedral' QPO, the best fit model, and the single components used in the fit. The solid 
lines show the thin features (QPO and harmonic), the dashed line show the continuum component (including the white noise). }
\plotone{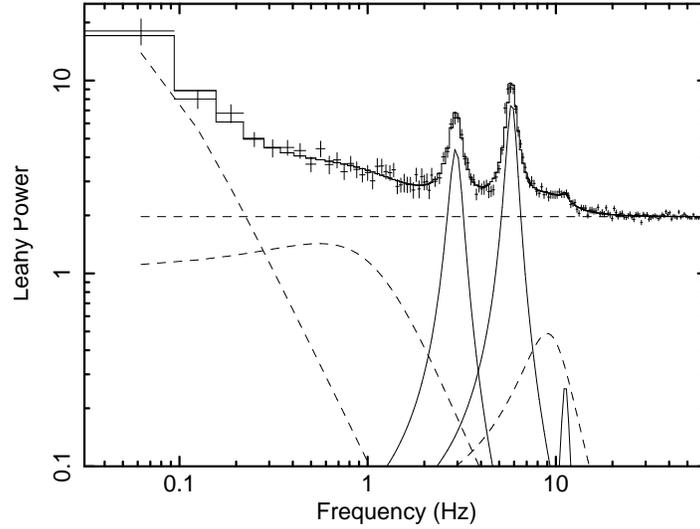}
\label{fig:PDS}
\end{figure}

We presented here an  analysis of some of the properties of the peculiar 'Cathedral' QPO \citep{casella04} seen in \X1859. Although  we mainly focused 
here on Obs. 1 to illustrate our analysis, similar results and trends were obtained in the case of Obs. 2, allowing us to use these two 
observations in our argumentation. The cathedral QPO manifests as two apparently harmonically-related peaks of similar amplitudes
 in the PDSs  (Fig. \ref{fig:PDS}).  \citet{casella04}  concluded that QPO$_2$ is the fundamental feature while QPO$_1$ is its 'sub'-harmonic. 
 With this conclusion, and adding the fact that QPO$_2$ shows hard time lags while QPO$_1$ show soft time lags, they classify 
 the Cathedrals as type B QPOs \citep{casella04}. \\

\begin{figure}
\epsscale{0.5}
\caption{Evolution 
of the QPOs absolute amplitudes with the count rate for Obs. 1. The diamonds correspond to QPO$_1$ and the  circles to QPO$_2$.}
\plotone{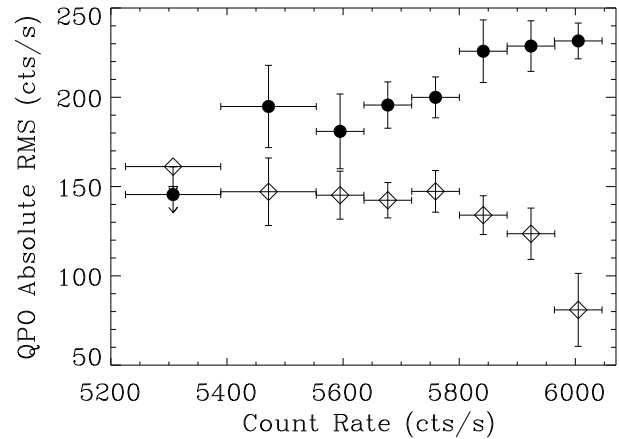}
\label{fig:QPOvsCount}
\end{figure}

\begin{figure}
\epsscale{1}
\caption{ {\bf{Left: }} Energy dependences of the two main QPOs. Upper limits are given at the 90\% level. {\bf{Right: }} Ratio between the amplitudes 
of QPO$_2$ and QPO$_1$.}
\plotone{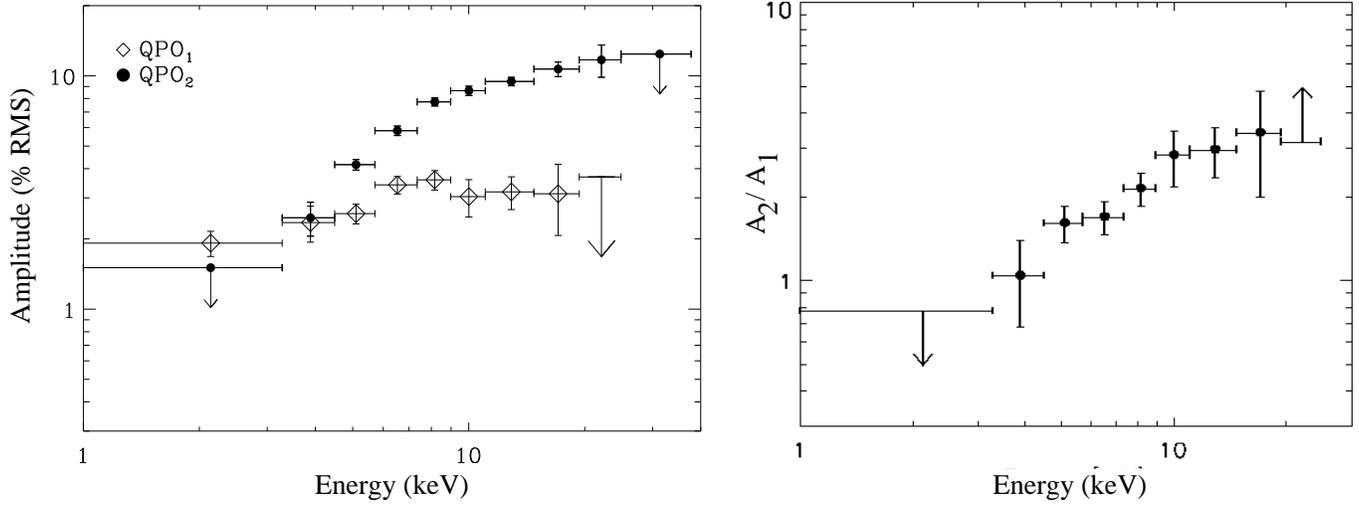}
\label{fig:QPOspec}
\end{figure}

\begin{figure}
\epsscale{0.5}
\caption{Bicoherence plot for Obs. 1 on a frequency region including the two peaks. The white regions are  $5\times5$ pixels boxes 
centered on the positions of the two peaks (at respectively $\sim(2.94,2.94)$~Hz and $\sim(5.83,5.83)$~Hz), while the black one is centered on 
$\sim(5.83,2.94)$~Hz and shows the value of $b^2$ at the intersection of the two peaks.}
\plotone{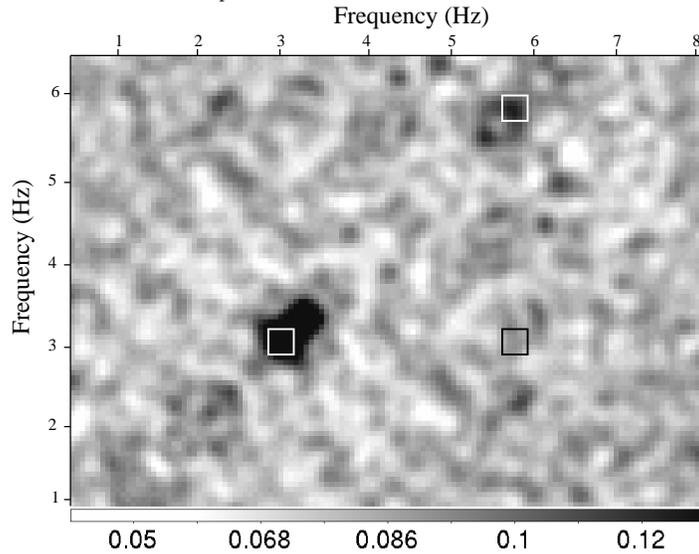}
\label{fig:bicoh}
\end{figure}

\indent Although QPO$_1$ and QPO$_2$ seems to be harmonically related, and have been classified as harmonics, 
 their overall behavior is quite different. We summarize here the main differences:
 \begin{enumerate}
 \item Their amplitude has a different temporal evolution. QPO$_2$ is  present 
 during the whole observation (apart from the 3 dips mentioned in Sec.~\ref{sec:dynpo}), while QPO$_1$ 
 is only intermittently seen. Our analysis shows that  it undergoes larger variations of its amplitude than QPO$_2$ (Sec. \ref{sec:dynpo}). 
 \item  The RMS-spectra of these two peaks are clearly different in shape, normalization and  typical parameters (Fig.~\ref{fig:QPOspec}). 
 In particular, the characteristic energy at which their spectra flatten is different by  about a factor of $\geq3$: QPO$_1$ reaches 
 its peak at around  $\sim6$ keV, while QPO$_2$ is much harder and peaks at $\gtrsim20$ keV. 
 \item The time lags of the two peaks are different: QPO$_1$ has soft lags (the soft X-rays lag behind the 
 hard ones), while QPO$_2$ has hard lag \citep{casella04}. This property is a definition of type-B QPOs \citep[e.g.][]{remillard02,casella05}.
 \itemÊThe value of the bicoherence $b^2(\nu_2,\nu_1)$ is compatible with statistical noise, which tends to indicate a very weak 
 or no coupling between the  peaks at $\nu_2$, $\nu_1$, and $\nu_1+\nu_2$.
 \end{enumerate}
 \indent While points 2, and 3 have been mentioned for other BHBs  \citep[e.g.][for XTE J1550$-$564 and  GRS 1915+105] {homan01,rao10,cui99}, 
 in addition to \X1859\ \citep{casella04}, it is the first time to our knowledge that points 1 and 4 are  reported for any BHBs. \correc{In 
 XTE J1550$-$564 \citet{heil11} report a positive linear RMS-flux relation for  (type-C) QPOs with a frequency smaller than $\sim 4$~Hz, and a 
 negative RMS-flux trend at higher frequencies. Here the situation is opposite. QPO$_2$, that has the highest frequency, shows the linear 
 relation, and QPO$_1$ shows the negative trend (Fig.~\ref{fig:QPOvsCount}). This could be the signature of a new fundamental 
 difference between types B and C QPOs.} \\
 \indent  In XTE J1550$-$564, \citet{rao10} mentioned 
 that the difference of energy spectra of the two peaks, especially the fact that the fundamental has a higher amplitude at high energy can
 indicate a more sinusoidal signal at higher energies. In this respect, the fact that QPO$_1$ is stronger at low flux 
 could also indicate a more sinusoidal signal in the peaks. This explanation,  although simple and tempting, do not account for the 
 different signs of the time lags of the two peaks, since one would naively expect in the case of a non-purely sinusoidal signal that all 
 components undergo the same physical processes and thus show similar lags. In addition, \citet{rao10} come to the conclusion
 that the peak referred to as sub-harmonic in XTE J1550$-$564 could in fact be the true fundamental feature. In this case the "more sinusoidal" 
 explanation does not hold any more since the harmonic (and therefore the peak with the highest frequency) should then disappear first, and show 
 a softer spectra, which is clearly opposite to what is seen here.\\
 \indent  It is interesting to remark, here, that the bicoherence behavior of the two peaks 
 (at $(\nu_1,\nu_1)$ and $(\nu_2,\nu_2)$) is similar to that of the  type C QPOs of GRS~1915+105 \citep{maccarone11}, where $b^2$ reaches 
 a high value close to the frequencies of the peaks. In this latter source $b^2$ is high at the (bi-)position of the fundamental, and much lower for the 
 harmonic \citep{maccarone11}.  Pursuing the comparison with GRS 1915+105, in \X1859\ the values of $b^2$ would tend to 
 indicate that QPO$_1$ is the fundamental and QPO$_2$ its harmonic.   In this latter source, however, different global patterns have been identified in 
 the bicohrence plot. In all cases (where a strong harmonic is seen)  a strong coupling is seen between 
the fundamental and the harmonic \citep[e.g. the "web" or "cross" patterns][]{maccarone11}, and even between the noise component and the 
QPO. This is  not observe in the case of \X1859\  (Fig.~ref{fig:bicoh}). \\
 \indent An easy interpretation for the four points above would be that the two peaks  
  have absolutely no relation together and that the apparent harmonic relation of their frequencies is fortuitous. This seems difficult to reconcile 
  with the fact that a similar behavior is seen both in Obs. 1 and 2 that are separated by roughly a day, and a transition into another state \citep{casella04}.
In addition, point 2 in the case of  type C QPOs, and points 2 and 3 in the case of type B QPOs  have been reported in this and other sources 
\citep{cui99, homan01, remillard02, rodrigue02_qpo, casella04, rao10}. It is more likely that the two peaks are somehow related or that, at least, 
a possible common mechanism sets their frequencies to be integer multiple, with the two features not being harmonics in a physical sense.\\
 \indent Hard lags are usually easy to understand in the context of Comptonization of soft photons in a hot and tenuous medium \citep[e.g. the so-called
 corona; e.g. ][]{cui99}. As discussed in \citet{cui99}, and \citet{gierlinski05} the energy dependences of both the QPO and/or the 
 continuum are indicative of variations of physical parameters: the favored ones are either the temperature and/or optical depth of the corona \citep{cui99},
 the temperature of the soft photons,  or a variable power of the Comptonized component \citep{gierlinski05}. In this respect, however, the different 
 time dependences of the two peaks, and the opposite sign  of their time lags are difficult to understand. \\
 \indent Other families of models involve precession at the inner boundary of the disk, but here again, the fundamental and harmonics 
 are produced at the same location and should 'see' the same environment. In the framework of Comptonization the time lags,  
 RMS-spectra, and bicoherence  of the different features should be the same. \\
 \indent A way to possibly reconcile the four points summarized above could be that the peaks are not harmonics in the usual sense 
 but that they represent different modes of the same mechanism favored at different moments, depending of some (unknown) parameter(s)
 in the corona-disk-jet system. This hypothesis has the advantage of providing an explanation for the time behavior of the two peaks, and 
 permits in particular that they do not necessarily appear at the same time, but enter in a sort of competition.  The competition has also been 
 mentioned for other types of QPOs in \X1859 by \citet{casella04} and in the case of XTE J1550$-$564 by \citet{remillard02} although it has never 
 been explored further.   The physical states of the 
 disk/corona/jet system might then set the conditions for one or the other peak to dominate. This will be explored by us from a particular theoretical 
standpoint in a forthcoming paper (Varni\`ere, Tagger, Rodriguez, submitted to A\&A). we note that the behavior observed in the case of 
the cathedral QPO is different from, at least, that of the type C QPO of GRS~1915+105 with respect to the \correc{RMS-flux relation and to the} 
bicoherence behavior \citep{heil11,maccarone11}. This may, however,  represent a fundamental property of type B QPOs in general. 
In any cases, the new model-independent  observational facts presented here should be taken as strong constraints in any 
attempts to model LFQPOs in BHBs.\\
 
\begin{acknowledgements}
The authors warmly thank I. Caballero, S. Corbel, and M. Cadolle Bel  for a careful reading of early versions of the paper 
and useful comments. We also warmly thank T. Maccarone and P. Uttley for very fruitful discussions about the  
interpretation of the bicoherence, and for indicating us a small mistake in the first version of the paper. 
We acknowledge the referee for his/her very useful and helpful report that really helped the paper to be improved.   
J.R. acknowledges partial funding from the European CommunityÕs Seventh Framework
Programme (FP7/2007-2013) under grant agreement number ITN
215212 "Black Hole Universe". This work has been financially partially supported by the GdR PCHE in France. 
This research has made use of data obtained through the High Energy 
Astrophysics Science Archive Center Online Service, provided by the NASA/Goddard Space Flight Center.
\end{acknowledgements}


\begin{table}
\caption{QPO parameters obtained from the fits to the 2--40 keV PDS. The errors are given at the 90\% level. }
\begin{tabular}{l c c c c c c c c c}
 Obs. \# &$\nu_1$ & Q$_1$$^\ddagger$ & $A_1$$^\dagger$ & $\nu_{2}$ & Q$_{2}$$^\ddagger$ & $A_{2}$$^\dagger$ & $\nu_{3}$ & Q$_{3}$$^\ddagger$ & $A_{3}$$^\dagger$ \\
 & (Hz)       &                                      & (\% RMS) & (Hz)       &                                      & (\% RMS) & (Hz)       &                                      & (\% RMS) \\
\hline
1 &  $2.94\pm0.02$ & $5.9$ & $2.8\pm0.1$ & $5.828\pm0.025$ & $7.3$ &$4.7\pm0.1$ & $11.2_{-0.4}^{+0.3}$& 9.5 & $1.1_{-0.3}^{+0.2}$\\
2 &  $3.00\pm0.04$ & $5.2$ & $2.9\pm0.2$ & $5.86\pm0.04$ & $6.5$ &$4.6_{-0.1}^{+0.2}$ & $11.5\pm0.3$& 8.3 & $1.5_{-0.3}^{+0.2}$\\
\hline
\end{tabular}
\begin{list}{}{}
\item[$^\ddagger$]Q=$\nu/{\mathrm{FWHM}}$.
\item[$^\dagger$]$A$ stands for RMS amplitude.
\end{list}
\label{tab:qpofit}
\end{table}

\end{document}